\documentclass[namedreferences,letterpaper]{SolarPhysics}

\usepackage[optionalrh,solaenum]{spr-sola-addons} 
\usepackage{graphicx}
\usepackage{courier}
\usepackage{amssymb}
\usepackage{color}
\usepackage{url}

\def\arcsec{\hbox{$^{\prime\prime}$}}

\begin{document}

\begin{article}

\begin{opening}

\title{On-Orbit Sensitivity Evolution of the {\textit{\textbf EUV
      Imaging Spectrometer}} on {\textit {\textbf Hinode}}}

\author{J.T.~\surname{Mariska}$^{1}$
       }

%
\runningauthor{J.T. Mariska}
\runningtitle{\textit{Hinode} EIS Sensitivity Evolution}

%
  \institute{$^{1}$ School of Physics, Astronomy, and
    Computational Sciences, George Mason University, 4400
    University Drive, Fairfax, VA 22030, USA
                     email: \url{jmariska@gmu.edu}
             }

\begin{abstract}
Since its launch on 22 September 2006, the \textit{EUV Imaging
  Spectrometer} onboard the \textit{Hinode} satellite has
exhibited a gradual decay in sensitivity. Using spectroheliograms
taken in the Fe~{\sc viii} 185.21~\AA\ and Si~{\sc vii}
275.35~\AA\ emission lines in quiet regions near Sun center we
characterize that decay. For the period from December 2006 to
March 2012, the decline in the sensitivity can be characterized
as an exponential decay with an average time constant of $7358
\pm 1030$ days ($20.2 \pm 2.8$ years). Emission lines formed at
temperatures $\gsim\! 10^{6.1}$ K in the quiet-Sun data exhibit
solar-cycle effects.
\end{abstract}

%

\end{opening}


\section{Introduction}\label{s:introduction}

The \textit{EUV Imaging Spectrometer} (EIS) on the
\textit{Hinode} satellite \cite{Kosugi2007ex} observes emission
lines of highly ionized elements in two wavelength bands with the
aim of using line intensity, profile, and Doppler shift data to
characterize plasma properties in the solar atmosphere at
temperatures ranging from $\approx\! 40,000$~K to $\approx\!
20$~MK \cite{Culhane2007ex}. All of the individual components of
EIS were characterized before final instrument assembly, and the
instrument underwent end-to-end testing and calibration
\cite{Korendyke2006ex,Lang2006ex}. These data were used to
compute effective area curves for each EIS wavelength band.

After end-to-end testing, EIS was shipped to Japan in August
2004, stored in a controlled environment at the Japanese
Aerospace Exploration Agency, integrated onto the \textit{Hinode}
satellite, and launched on 22 September 2006. During this time
interval, it was not possible to monitor the status of the
instrument calibration. On orbit, EIS was allowed to outgas for
30~days, and the first spectra were acquired on 28 October 2006.

All previous solar EUV spectrometers have exhibited sensitivity
degradation over time. Thus, early in the mission, the EIS team
initiated a program of regular quiet-Sun observations of selected
EUV lines to monitor the instrument sensitivity. This article
reports on an analysis of those data.

\section{Observations and Data Processing}\label{s:observations}

EIS observes the solar EUV spectrum in two
40-\AA\ wavelength bands centered at 195 and 270~\AA. Each
spectrum is stigmatic along the slit, which is oriented in the
N\,--\,S direction. The instrument produces line profiles by
imaging the Sun onto two slits ($1\arcsec$ and $2\arcsec$), and
monochromatic images by imaging the Sun onto two slots
($40\arcsec$ and $266\arcsec$). Moving a fine mirror mechanism
allows EIS to construct spectroheliograms in selected emission
lines by rastering regions of interest. \inlinecite{Culhane2007ex}
provide an extensive discussion of EIS and its operation.

From 6 April 2007 through 28 January 2008, EIS regularly obtained
spectroheliograms covering $128\arcsec \times 184\arcsec$ with a
step size of $1\arcsec$ in quiet-Sun regions generally near
Sun-center (EIS study SYNOP002). Each 90-second exposure covered
the entire wavelength range of both CCDs. From each exposure, 14
wavelength windows containing relatively strong lines and
covering the wavelength ranges of both detectors were extracted
for further analysis.

Early in 2008, the \textit{Hinode} X-band transmitter failed and
data transmission from the satellite had to rely on the slower
S-band transmitter. In response to the reduced data rates the EIS
team shifted to a new monitoring study (SYNOP006), which had the
same raster size and exposure time but only retrieved data from
the 14 wavelength windows that were being analyzed using data
from the earlier study. Aside from the change to transmitting
just the 14 wavelength windows, the new study was identical to
the earlier one.

Since it is possible that significant sensitivity loss occurred
early in the mission, the studies outlined above were
supplemented by searching the EIS data for similar
spectroheliograms taken near Sun center from early in the mission
until April 2007, when the first synoptic study was initiated.
These data usually did not contain all of the lines captured in
the synoptic studies, but they provide significant useful
information on the instrument sensitivity changes early in the
mission.

All of the data sets used in this study were processed using the
standard EIS data reduction software. This software removes
detector bias and dark current, hot pixels, dusty pixels, and
cosmic rays, and then applies the prelaunch absolute calibration.
This results in sets of intensities in erg cm$^{-2}$ s$^{-1}$
sr$^{-1}$ \AA$^{-1}$. All of the emission lines, with the
exception of He~{\sc ii} 256.317~\AA, are isolated features.
Thus, in principle, it is possible to compute the total intensity
in each line by simply summing the pixel values with the
background subtracted. This approach, however, does not provide
any simple measure of which profiles are unacceptable because,
for example, of improperly removed artifacts or cosmic ray hits.
Instead, for each selected wavelength window, the emission line
data were fitted with Gaussian line profiles plus a constant
background. This has the advantage of providing an indication of
the goodness of the fit, which can be used to exclude bad data.
The resulting total line-intensity values are the basic data used
in this analysis. Table~\ref{tbl:lines} lists the emission lines
used in this study along with their temperature of formation. The
additional columns in the table will be discussed later in this
article.

\begin{table}
\caption{Emission lines used in this study and estimated
  intensities on 6 November 2007. The errors listed for the EIS
  intensities were computed by taking the standard deviation of
  the residuals between the model for the EIS sensitivity decline
  and the observed data.}\label{tbl:lines}
\begin{tabular}{lcccc}
\hline
Ion & $\lambda$ & $\log T_{\mathrm e}$ & $I$ & $I_{\mathrm{cal}}$\\
& [\AA] & [K] & \multicolumn{2}{c}{[erg cm$^{-2}$ s$^{-1}$ sr$^{-1}$]}\\
\hline
He~{\sc ii} & 256.317 & 4.85 & $169\pm 19$ & \ldots\\
Fe~{\sc viii} & 185.213 & 5.65 & \phantom{0}$27\pm 5.5$ &
\phantom{0}$36\pm 3.6$\\
Si~{\sc vii} & 275.352 & 5.80 & \phantom{0}$14\pm 2.9$ & \ldots\\
Fe~{\sc x} & 184.536 & 6.00 & \phantom{0}$76\pm 17$ & $122\pm 12$\\
Fe~{\sc xi} & 180.401 & 6.10 & $199\pm 52$ & $345\pm 34$\\
Fe~{\sc xii} & 193.509 & 6.15 & \phantom{0}$64\pm 35$ &
\phantom{0}$95\pm 9.5$\\
Fe~{\sc xii} & 195.119 & 6.15 & \phantom{0}$89 \pm 47$ & \ldots\\
Si~{\sc x} & 258.375 & 6.15 & \phantom{0}$31\pm 9.6$ & \ldots\\
Fe~{\sc xiii} & 202.044 & 6.20 & \phantom{0}$47\pm 47$ & \ldots\\
\end{tabular}
\end{table}

\begin{figure} 
\centerline{\includegraphics[width=1.0\textwidth]{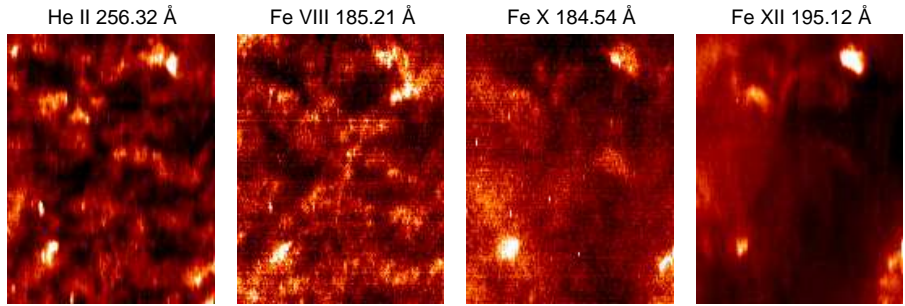}}
\caption{Spectroheliograms constructed from data taken starting
  at 14:33:20 on 7 January 2008. The images have dimensions of
  128\arcsec\ in the $x$-direction and 166\arcsec\ in the
  $y$-direction, and are centered at an ($x$,$y$) position on the
  Sun of ($-290\arcsec$, $577\arcsec$). The $y$-direction size is
  smaller than the true $y$-dimension of the data because some of
  the rows are lost when the offset between the two detectors is
  used to co-register the data.}\label{fig:rasters}
\end{figure}

Figure~\ref{fig:rasters} shows an example of some of the
intensity spectroheliograms obtained from one of the data sets
used in this work. The spectroheliograms exhibit the typical
features of the quiet Sun seen in lines formed in the transition
region and corona. In the transition region, lines such as those
of He~{\sc ii} and Fe~{\sc viii} exhibit the well-known
cell-network pattern with some areas of network being
significantly brighter than the overall average. Coronal lines
such as those of Fe~{\sc x} and Fe~{\sc xii} show a smoother
emission pattern, but some of the brighter network emission
locations are also bright at the higher temperatures.

Each year from roughly late April until late August (the
beginning and ending dates are changing as the orbit changes),
\textit{Hinode} passes through the Earth's shadow each orbit.
Since a full synoptic study takes more than an orbit to complete,
some of the exposures will be compromised. These exposures were
removed from the data set by visually inspecting plots of the
total intensity in each exposure as a function of the solar
$x$-position in the Fe~{\sc xii} 195~\AA\ fitted intensity data.
Exposures with incomplete data were also removed.

The standard EIS data-reduction software includes an estimate of
the error for each pixel in an exposure, and these errors are
used by the fitting software to determine the reduced $\chi^2$
for the fit. This error estimate consists of the noise due to
photon statistics and an estimate of the dark-current uncertainty
combined in quadrature. For well-behaved line profiles the
average values of the reduced $\chi^2$ are generally around 0.5
rather than the expected value of 1.0. This effect was also noted
in CDS data by \inlinecite{Thompson2000}, who suggested a
renormalization procedure for the CDS error estimates. The
details of making adjustments in the EIS errors are still under
study. For this work, all intensities for which the reduced
$\chi^2$ for the fit is greater than 1.0 have been excluded.

The He~{\sc ii} 256.317~\AA\ emission line (actually a blend of
lines at 256.317~\AA\ and 256.318~\AA) is particularly
challenging to fit. On its long-wavelength side the line is
blended with several coronal lines. According to the CHIANTI
database \cite{Dere1997,Landi2012}, the lines are Si~{\sc X} at
256.378~\AA, Fe~{\sc x} at 256.398~\AA, Fe~{\sc xiii} at
256.400~\AA, and Fe~{\sc xii} at 256.410~\AA. In quiet solar
regions, the He~{\sc ii} line is significantly stronger than the
other lines and they can be represented by a second Gaussian,
which is mostly emission from the Si~{\sc x} line. Brighter
locations in the spectroheliograms often have somewhat stronger
emission from the hotter lines, and the fitting process then can
become challenging.

\begin{figure} 
\centerline{\includegraphics[width=1.0\textwidth]{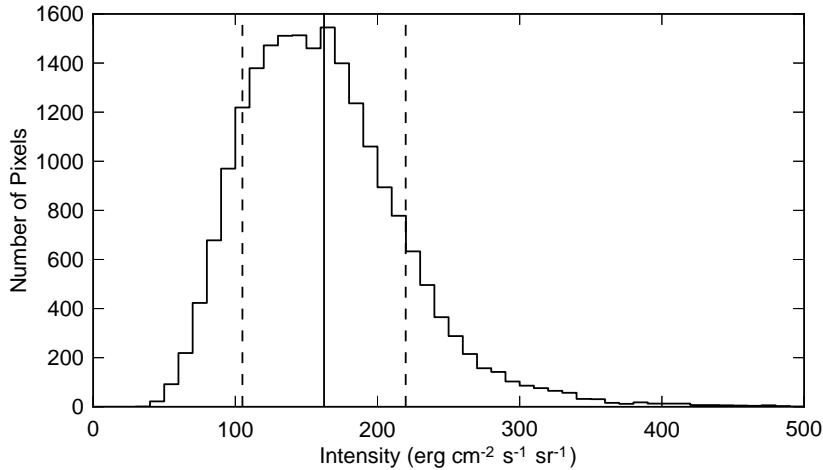}}
\caption{The distribution of He~{\sc ii} 256.317~\AA\ line
  intensities for the 7 January 2008 data shown in
  Figure~\ref{fig:rasters}.}\label{fig:sample-avg}
\end{figure}

For each spectroheliogram in each emission line, the line
intensities in the acceptable spatial pixels display a range of
intensities. As an example, Figure~\ref{fig:sample-avg} shows a
histogram of the distribution of He~{\sc ii}
256.317~\AA\ intensities in the data set shown in
Figure~\ref{fig:rasters}. The histogram displays the
characteristics typical of emission lines from the upper
chromosphere and lower transition region (\textit{e.g.}
\opencite{Reeves1976}; \opencite{Schrijver1985}). The number of
pixels at a given intensity rises rapidly to a broad peak and
then exhibits an extended higher-intensity tail compared with the
rising part of the histogram. This behavior is a characteristic
of a log-normal intensity distribution (\textit{e.g.}
\opencite{Warren2005a}). A solid line in the figure marks the
mean intensity for the spectroheliogram. Dashed lines show one
standard deviation around the mean. For this spectroheliogram,
the average intensity is 162 erg cm$^{-2}$ s$^{-1}$ sr$^{-1}$
with a standard deviation of 57.4 erg cm$^{-2}$ s$^{-1}$
sr$^{-1}$. This value for the standard deviation is typical for
most of the He~{\sc ii} spectroheliograms and is a reflection of
the range of intensities seen in the quiet Sun. The width of the
Gaussian characteristic of the log-normal distribution
represented by the data in Figure~\ref{fig:sample-avg} is 0.15,
comparable to the values tabulated in \inlinecite{Warren2005a}.
Histograms for other emission lines display similar behavior for
the cooler lines. For the hottest lines considered in this study,
those with a temperature of formation above that of Fe~{\sc xii},
the histograms often have more than one peak, showing the
influence of small regions of enhanced emission in the
spectroheliogram.

\section{Analysis}\label{s:analysis}

\begin{figure} 
\centerline{\includegraphics[width=1.0\textwidth]{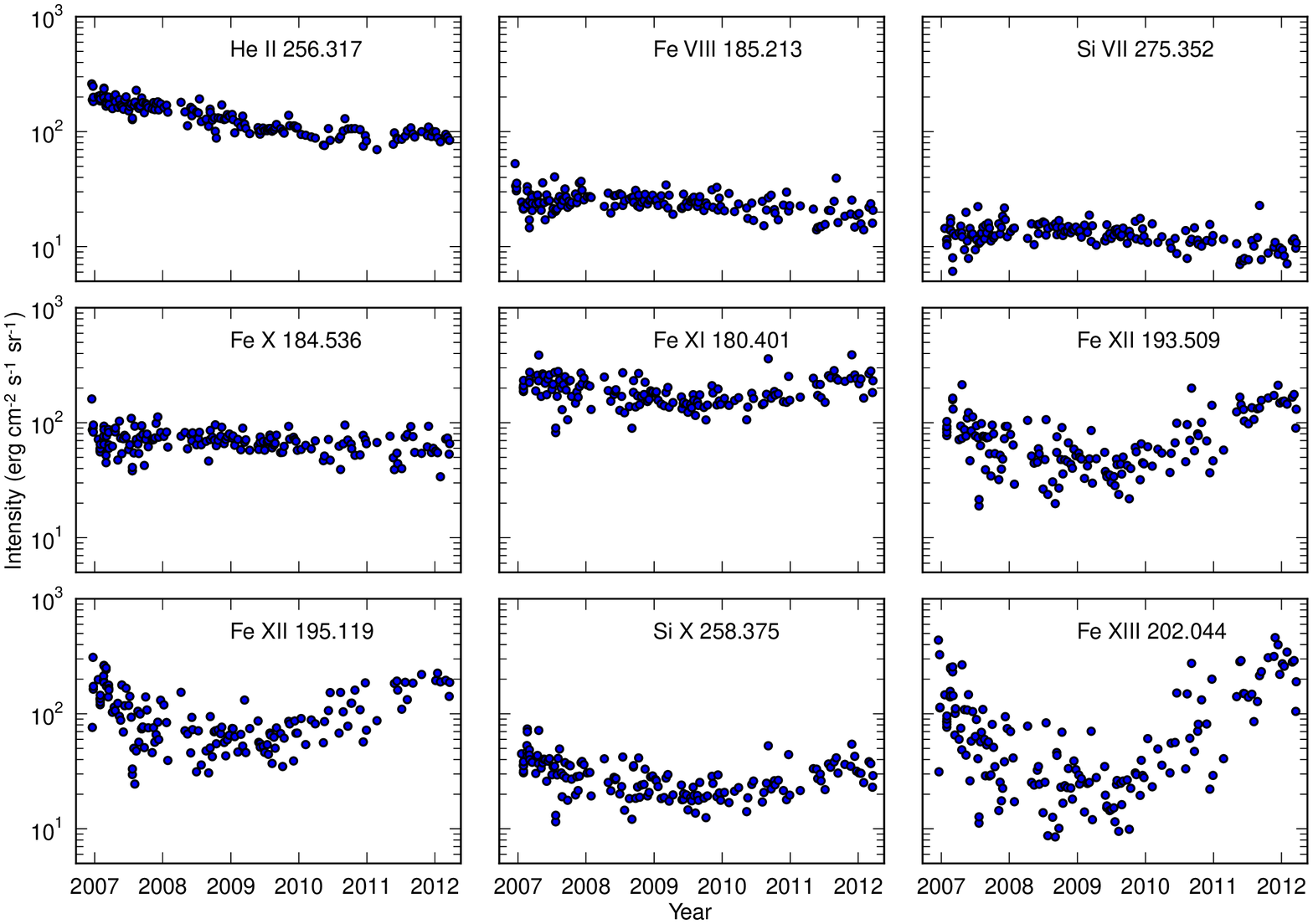}}
\caption{Average line intensities as a function of time for a
  selection of the emission lines observed in the EIS monitoring
  studies.}\label{fig:time-hist}
\end{figure}

For each data set and emission line listed in
Table~\ref{tbl:lines} average intensities and standard deviations
were computed in the manner outlined above.
Figure~\ref{fig:time-hist} displays the temporal behavior of the
averages for nine of the 14 emission lines captured in the
standard monitoring studies. The data in both the figure and in
Table~\ref{tbl:lines} are ordered by the temperature of formation
of the emission lines rather than by wavelength. To show better
the range of intensity variations among the emission lines, all
of the data have been plotted using the same range of values on
the $y$-axes. Averages for the higher-temperature emission lines,
those of Fe~{\sc xi}, Fe~{\sc xii}, Si~{\sc x}, and Fe~{\sc
  xiii} show an initial decline in average intensity as a
function of time, but then show an increasing average intensity
as a function of time after some time in 2009. The NOAA Space
Weather Prediction Center lists December 2008 as the date of the
recent solar minimum. Thus it appears that even in quiet solar
regions the hotter emission lines are affected by the solar
cycle. This effect has already been noted by
\inlinecite{Kamio2012}. Averages for the lower-temperature
emission lines in the figure, those of He~{\sc ii}, Fe~{\sc
  viii}, Si~{\sc vii}, and Fe~{\sc x}, show declining average
intensities as a function of time for the entire time period.

The slope of the initial decline in average intensities for the
higher temperature lines is generally steeper at the higher
temperatures than at the lower temperatures. For the emission
lines of Fe~{\sc viii}, Si~{\sc vii}, and Fe~{\sc x} the slopes
are similar, with the average intensities declining by about
25\,\% over the time period of the observations.

The average emission in the He~{\sc ii} 256~\AA\ emission line is
somewhat different. As we noted earlier, this line is challenging
to fit accurately. The He~{\sc ii} data shown in
Figure~\ref{fig:time-hist} are for only the
256.317~\AA\ component of the fit. These data show a somewhat
different behavior than the emission from the next three hottest
lines of Fe~{\sc viii}, Si~{\sc vii}, and Fe~{\sc x}. The initial
decline of the He~{\sc ii} emission is more rapid than those two
lines, but then the trend flattens and there is a much smaller
decrease in the last two years of data. The similarity of the
initial decline to the behavior of the hotter coronal lines
suggests that the two-component fits to the He~{\sc ii} data have
not been fully successful at separating the 256.317~\AA\ emission
line from the hotter emission lines.

\begin{figure}
\centerline{\includegraphics[width=0.5\textwidth]{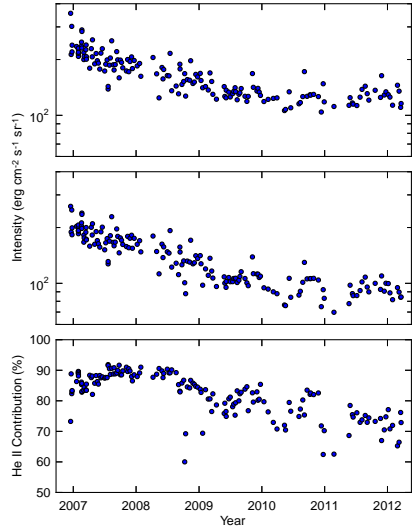}}
\caption{He~{\sc ii} 256.317~\AA\ line intensities as a function
  of time. The top panel is the total intensity obtained using a
  two-component fit to the data, the middle panel shows the
  He~{\sc ii} 256.317~\AA\ component only, and the bottom panel
  shows the percentage contribution of that component to the
  total intensity.}\label{fig:he-panel}
\end{figure}

Figure~\ref{fig:he-panel} shows the He~{\sc ii}
256.317~\AA\ data, the total intensity in the line profile, and
the percentage of the emission coming from the
256.317~\AA\ component as a function of time. Both the total
intensity and the 256.317~\AA\ component show the rapid decline.
As that decline takes place, the contribution of the He~{\sc ii}
component decreases. This is consistent with the solar-cycle
behavior exhibited by the coronal emission lines of Fe~{\sc xii},
Si~{\sc i}, and Fe~{\sc xiii}. In fact, if the He~{\sc ii}
256.317~\AA\ averaged intensities are assumed to be
characteristic of the decline in EIS sensitivity and are used to
detrend the other data, then the resulting intensities in the
Fe~{\sc viii}, Si~{\sc vii}, and Fe~{\sc x} increase throughout
the time period of the observations. Since the solar minimum did
not occur until the end of 2008, this behavior is not plausible.
We therefore conclude that, despite the fact that the He~{\sc ii}
emission line is formed at a lower temperature than any of the
other emission lines used in this study and should therefore be
less subject to solar-cycle effects, the uncertainties in
removing the coronal emission from the line intensities result in
a data set that is seriously compromised.

\begin{figure} 
\centerline{\includegraphics[width=0.5\textwidth]{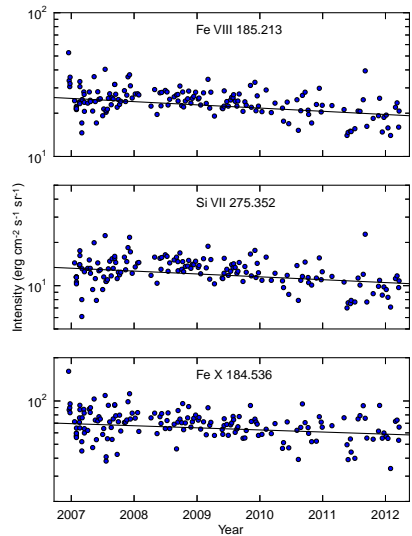}}
\caption{Average line intensities in the Fe~{\sc viii}, Si~{\sc
    vii}, and Fe~{\sc x} emission lines. The lines on each plot
  show the results of an exponential fit to each data
  set.}\label{fig:cool-fit-plots}
\end{figure}

Since the He~{\sc ii} data are compromised, the other emission
lines that show minimal obvious solar cycle effects offer the
most promise for estimating any EIS sensitivity loss.
Figure~\ref{fig:cool-fit-plots} shows the average intensities in
the Fe~{\sc viii}, Si~{\sc vii}, and Fe~{\sc x} emission lines
along with an exponential fit to the data. To perform these fits,
it is necessary to weight each data point. Each individual pixel
in each spectroheliogram has associated with it an error in the
total intensity. Since a large number of individual emission line
profile fits go into each averaged data point shown in the
figure, however, the formal error associated with the average is
very small. Two other weighting choices better capture the
variations seen in the data. One is to weight each averaged data
point by the number of individual measurements that went into its
determination. The other is to weight each point by the standard
deviation of the average of all of the data points in the
spectroheliogram. Each weighting results in nearly the same
results for the two parameters of the fit. The solid curves in
Figure~\ref{fig:cool-fit-plots} show the best fit to an
exponential decay using the standard deviations for the
weighting. The resulting $1/e$ decay times are $7088 \pm 1247$,
$7939 \pm 1830$, and $10\,870 \pm 3288$ days for the Fe~{\sc
  viii}, Si~{\sc vii}, and Fe~{\sc x} emission lines,
respectively. Given the very similar slopes of the fits to the
Fe~{\sc viii} and Si~{\sc vii} fits, we have combined the results
and compute a weighted mean value for the time constant of the
EIS sensitivity decay of $7358 \pm 1030$ days ($20.2 \pm 2.8$
years).

\section{Discussion}\label{s:discussion}

The Fe~{\sc viii} 185.213~\AA\ and Si~{\sc vii}
275.32~\AA\ emission lines are captured by separate EIS
detectors. Since the decay constants for the decline in emission
are similar, we conclude that the two detectors are exhibiting
similar sensitivity changes with time. Thus, it is likely that a
single decay constant can be used to define the EIS sensitivity
loss at all wavelengths. Also, this suggests that the gradual
decline may be due to a loss of throughput elsewhere in the EIS
optical system.

\begin{figure} 
\centerline{\includegraphics[width=1.0\textwidth]{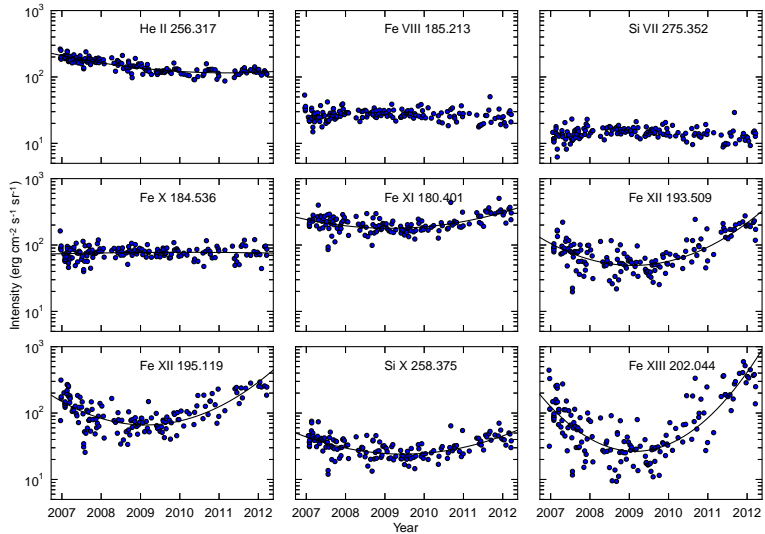}}
\caption{Average line intensities as a function of time for the
  data shown in Figure~\ref{fig:time-hist} detrended using the
  fit to the Fe~{\sc viii} 185.213~\AA\ and Si~{\sc vii}
  275.352~\AA\ intensity decays.}\label{fig:detrended-data}
\end{figure}

As we pointed out in Section~\ref{s:analysis}, the emission lines
formed at and above the temperature of formation of the Fe~{\sc
  xi} 180.401~\AA\ line exhibit an apparent solar-cycle effect.
Assuming that the Fe~{\sc viii} and Si~{\sc vii} line intensity
changes are purely instrumental, we can use the decay constant
obtained above to detrend all of the emission line data shown in
Figure~\ref{fig:time-hist}. Figure~\ref{fig:detrended-data} shows
the results of that detrending. The detrended Fe~{\sc viii} and
Si~{\sc vii} data are included in the figure and, as expected,
are nearly constant as a function of time. As was the case with
the non-detrended data, the hotter emission lines from Fe~{\sc
  xiii}, Si~{\sc x}, Fe~{\sc xii}, and Fe~{\sc xi} continue to
exhibit a solar-cycle-related trend. The He~{\sc ii} intensities
continue to show the behavior discussed earlier\,--\,emission in
a low transition-region line contaminated by a coronal component.

Following the approach used in \inlinecite{Kamio2012}, we fit the
detrended data with a quadratic function to show better the
behavior as a function of time. Solid lines on each panel in
Figure~\ref{fig:detrended-data} show the results of that fitting.
Using a different but very similar data set,
\inlinecite{Kamio2012} found that the hotter emission lines
showed a minimum in February 2009. Our data, detrended using the
newer fit obtained above, show a similar behavior. The emission
lines of Fe~{\sc xiii}, Fe~{\sc xii}, and Fe~{\sc xi} all show a
minimum in February 2009. For the Si~{\sc x} line, the fit has a
minimum in June 2009. Note that both the cycle minimum date
provided by the Space Weather Prediction Center and the results
derived in \inlinecite{Kamio2012} were obtained using data that
were smoothed in time.

It is also challenging to determine how the overall solar cycle
measured in the 10.7-cm radio flux or the sunspot number relates
to measurements of a small region of quiet Sun. Since the
activity manifestations of the solar cycle begin at high
latitudes and gradually migrate toward the Equator, we would
expect their effect on small regions of nominally quiet Sun near
Sun center to vary. Early in the \textit{Hinode} mission, the Sun
was declining from its maximum in 2002. Thus much of the activity
was near the Equator and we might expect some influence on EIS
spectroheliograms obtained there, even for quiet regions.
Activity from the new cycle, however, was initially at high
latitudes. Thus, we would expect it to have a smaller impact on
EIS quiet-region observations near the Equator. That impact
should then grow over time.

\inlinecite{Kamio2012} detrended their data set using an earlier
sensitivity analysis that relied on the He~{\sc ii}
256.317~\AA\ data that we have rejected as unsuitable. While this
analysis alters the details of the time histories of the emission
lines they studied, it does not alter their primary conclusion
that the high-temperature component of the quiet corona changes
with time, suggesting that the heat input to the quiet corona
varies with time. This study further supports that conclusion.

The nature of the high-temperature emission component that
overlies quiet-Sun regions and fluctuates as the solar cycle
evolves has been the subject of considerable study.
\inlinecite{Vaiana1973} noted that at soft X-ray wavelengths the
quiet corona consisted of large-scale structures connecting
regions of opposite polarity. As the solar cycle evolves, the
structures necessarily evolve as a result of the changing
photospheric magnetic structure. Examining \textit{Yohkoh/Soft
  X-Ray Telescope} observations \inlinecite{Hara1997} noted that
the quiet-Sun soft X-ray flux decreased with the decreasing
magnetic flux as the solar cycle declines. \inlinecite{Acton1999}
have further quantified the behavior of the soft X-ray emission
observed with that instrument. \inlinecite{Orlando2001} examined
full-disk \textit{Yohkoh/Soft X-Ray Telescope} images as the
solar cycle declined from 1992 to 1996. The images they present
show a clear change from significant amounts of the so-called
quiet corona consisting of large-scale structures near the
maximum of the solar cycle to a quiet corona with significantly
less large-scale overlying structure near the cycle minimum.
\inlinecite{Kamio2012} discuss further the possible energetic
implications of these structural changes.

We have assumed in this analysis that the solar cycle does not
affect the emission from the Fe~{\sc viii} and Si~{\sc vii} lines
used to estimate the sensitivity decay constant. The lack of any
significant upturn in the averaged intensities in these lines as
the solar cycle rises suggests that this is the case. Continued
operation of EIS through the upcoming solar maximum and beyond
should help clarify the relative importance of the solar cycle in
these data. If the solar cycle is lifting the intensities in the
cooler lines, then the decay constant that we have derived will
be an upper limit to the actual sensitivity decline. Additional
insight into this issue might also be provided by examining
quiet-Sun spectra in transition-region lines observed throughout
the solar cycle with the \textit{Solar Ultraviolet Measurements
  of Emitted Radiation} on the SOHO satellite \cite{Wilhelm1995}.
Such a study, however, is beyond the scope of this article.

It is generally believed that volatiles condensing on the optical
surfaces and CCDs of EIS are the most likely source of
sensitivity loss (C.M. Brown, private communication, 2012). The
primary absorber is then carbon. In the EIS wavelength range, a
thin layer of carbon has a declining transmission as a function
of wavelength \cite{Henke1993}. Thus we would expect the
long-wavelength EIS channel to show a greater loss of sensitivity
with time than the short-wavelength channel. To within the errors
of the fits to the Fe~{\sc viii} and Si~{\sc vii} lines, that is
not the case. The difference in transmission of a 1000~\AA\ thick
layer of carbon between 185 and 275~\AA, however, is only about
15\,\%. Moreover, the two lines are formed at slightly different
temperatures, and thus solar cycle effects could affect them
differently. It may simply be that the wavelength dependence of
the absorbers is masked by these effects. At some point in the
future EIS will probably perform a bakeout of the CCDs to
ameliorate the effect of warm and hot pixels. If that also
significantly alters the sensitivity, it will add support to the
idea that volatiles are the source of the sensitivity loss.
Further analysis of data from well-studied ions with emission
lines on both CCDs should also help to clarify this issue.

While the primary focus of this article is the decline in the
sensitivity of EIS over time, it is useful to ask how the
absolute intensities compare with other observations.
\inlinecite{Wang2011} have reported the results of a comparison
of observations taken with a well-calibrated rocket experiment
flown on 6 November 2007 and near-simultaneous EIS observations.
The last two columns in Table~\ref{tbl:lines} list the
intensities of the emission lines used in this study on the date
of the rocket flight and the calibration-rocket results for the
lines observed by \inlinecite{Wang2011}. The errors listed for
the EIS intensities were computed by taking the standard
deviation of the residuals between the model for the EIS
sensitivity decline and the observed data. All of the intensities
deduced using the results of this work are lower than those
obtained by \inlinecite{Wang2011}. Given the very long
time-constant for the EIS sensitivity decline, this suggests that
the EIS prelaunch calibration may need an adjustment. A second
calibration rocket in the near future may provide valuable
additional insight.

Many earlier spaceborne EUV spectroscopic experiments exhibited
rapid sensitivity declines. For example, the \textit{Harvard
  College Observatory Spectroheliometer} on \textit{Skylab}
showed a loss of sensitivity at 977~\AA\ of $1/e$ over a 250 day
period \cite{Reeves1977}. The decline in sensitivity determined
in this study, however, is consistent with that seen in more
recent experiments such as the \textit{Coronal Diagnostic
  Spectrometer}, which saw an overall decline in sensitivity of
only about a factor of two over a 13-year period
\cite{DelZanna2010}. This implies a $1/e$ time of almost 19
years.

\begin{acks}
\textit{Hinode} is a Japanese mission developed, launched, and
operated by ISAS/JAXA in partnership with NAOJ, NASA, and STFC
(UK). Additional operational support is provided by ESA and NSC
(Norway). CHIANTI is a collaborative project involving the NRL
(USA), the Universities of Florence (Italy) and Cambridge (UK),
and George Mason University (USA). The author acknowledges
support from the NASA \textit{Hinode} program.
\end{acks}

%
%
\bibliographystyle{spr-mp-sola-cnd} 
\bibliography{allrefs}  

\end{article} 
\end{document}